\documentclass{aastex63}

\shorttitle{EUV Waves in a RMHD Simulation}
\shortauthors{Wang et al.}

\begin{document}

\title{Exploring the Nature of EUV Waves in a Radiative Magnetohydrodynamic Simulation}

\correspondingauthor{Feng Chen}
\email{chenfeng@nju.edu.cn}

\author{Can Wang}
\affiliation{School of Astronomy and Space Science, Nanjing University, Nanjing 210023, China}
\affiliation{Key Laboratory for Modern Astronomy and Astrophysics (Nanjing University), Ministry of Education, Nanjing 210023, China}

\author[0000-0002-1963-5319]{Feng Chen}
\affiliation{School of Astronomy and Space Science, Nanjing University, Nanjing 210023, China}
\affiliation{Key Laboratory for Modern Astronomy and Astrophysics (Nanjing University), Ministry of Education, Nanjing 210023, China}

\author{Mingde Ding}
\affiliation{School of Astronomy and Space Science, Nanjing University, Nanjing 210023, China}
\affiliation{Key Laboratory for Modern Astronomy and Astrophysics (Nanjing University), Ministry of Education, Nanjing 210023, China}

\begin{abstract}
Coronal extreme-ultraviolet (EUV) waves are large-scale disturbances propagating in the corona, whose physical nature and origin have been discussed for decades. We report the first three dimensional (3D) radiative magneto-hydrodynamic (RMHD) simulation of a coronal EUV wave and the accompanying quasi-periodic wave trains. The numerical experiment is conducted with the MURaM code and simulates the formation of solar active regions through magnetic flux emergence from the convection zone to the corona. The coronal EUV wave is driven by the eruption of a magnetic flux rope that also gives rise to a C-class flare. It propagates in a semi-circular shape with an initial speed ranging from about 550 to 700\,km s$^{-1}$, which corresponds to an average Mach number (\deleted{in }relative to fast magnetoacoustic waves) of about 1.2. Furthermore, the abrupt increase of the plasma density, pressure and tangential magnetic field at the wavefront confirms fast-mode shock nature of the coronal EUV wave. Quasi-periodic wave trains with a period of about 30\,s are found as multiple secondary wavefronts propagating behind the leading wavefront and ahead of the erupting magnetic flux rope. We also note that the true wavefront in the 3D space can be very inhomogeneous, however, the line-of-sight integration of EUV emission significantly smoothes the sharp structures in 3D and leads to a more diffuse wavefront.
\end{abstract}

\keywords{Radiative MHD, Solar EUV emission, Solar corona, Solar coronal waves, Solar activity}

\section{Introduction} \label{sec:intro}
Large-scale disturbances propagating in the corona, known as `EIT waves' or `EUV waves', were firstly detected by \citet{1998GeoRL..25.2465T} with the Extreme-ultraviolet Imaging Telescope (EIT) on the Solar and Heliospheric Observatory (SOHO) . Upon discovery, they were suggested to be the coronal counterpart of Moreton waves, which are chromospheric disturbances associated with solar flares  \citep{1960AJ.....65U.494M,1961ApJ...133..935A}. This intriguing phenomenon was extensively studied in the following decades with observations by EIT and the Solar Terrestrial Relations Observatory (STEREO) that provide a multi-perspective view. The Coronal EUV waves propagate in a wide range of speed spanning from a few tens to a few hundreds of km s$^{-1}$ \citep{2000A&AS..141..357K,2001A&A...370..591R,2014SoPh..289.4563M,2011ApJ...736L..13L} and exhibit various interactions with surrounding magnetic structures, for example, they can be refracted or reflected by the strong magnetic field, or transmit  cross the topological boundary of solar active regions and coronal holes \citep{2013ApJ...773L..33S,2009ApJ...691L.123G,2019ApJ...873...22S,2012ApJ...756..143O}. The complex behaviors revealed by observations led to a hot debate on the physical nature of EUV waves, and a variety of theoretical models have been put forward. The interpretation of the EUV waves can be divided into three major categories: wave model \citep[e.g., ][]{1974SoPh...39..431U,2004A&A...418.1101W,2007ApJ...664..556W} that considers EUV waves as fast magnetoacoustic waves or shocks, pseudo-wave model \citep[e.g., ][]{2007ApJ...656L.101A,2008SoPh..247..123D} that interprets the observed disturbance as a fact of reconfiguration of coronal magnetic field, and hybrid model \citep[e.g., ][]{2002ApJ...572L..99C,2005ApJ...622.1202C} that comprises of both wave and non-wave components. Comprehensive reviews on the observational properties and models of coronal EUV waves have been given by \citet{2014SoPh..289.3233L,2015LRSP...12....3W}, and \citet{2017SoPh..292....7L}.

In the past decade, the Atmospheric Imaging Assembly (AIA) on board Solar Dynamics Observatory (SDO) \replaced{allows people}{allowed} to acquire observation data with unprecedented high spatial and temporal resolutions, revealing more detailed features of EUV waves. Events with both wave and non-wave components are found to be very common \citep{2011ApJ...732L..20C,2012ApJ...745L..18A,2013ApJ...773..166L,2018A&A...612A.100C,2019SoPh..294...56F}. Such observations are also supported by three-dimensional (3D) models \citep{2009ApJ...705..587C,2012ApJ...750..134D}.

In spite of the continued improvement of observing techniques, the observational data have an obvious shortcoming that they cannot directly reveal the in situ physical parameters that are crucial to determine the nature of EUV waves. On the other hand, radiative magneto-hydrodynamic (RMHD) simulations with sophisticated physical processes allow for a direct and quantitative comparison between model synthesized observables and real solar observations. \added{Furthermore, although the physical processes considered in the simulation might not pose a significant direct impact on transient phenomena (such as coronal EUV waves), they play a crucial role in reproducing realistic thermodynamic properties of plasma, and this allows investigation of wave properties under conditions more similar to that of the real Sun.} Recently, realistic RMHD simulations have been applied to the production of solar flares for the first time \citep{2019NatAs...3..160C} and have successfully reproduced key properties of real solar flares. We present in this paper \added{the analysis of }the first realistic RMHD simulation of a solar flare that drives a coronal EUV wave, as well as the accompanying quasi-periodic wave trains. This provides us an exceptional dataset to study the evolution of physical quantities beneath the observational properties of the coronal EUV wave.

The rest of the paper is organized as follows. Section \ref{sec:method} gives a brief description on the numerical simulation. We present in Section \ref{sec:result} the analysis on the morphology and kinematics of the wave and changes of physical quantities across the wavefront. In the end, a discussion on the comparison between the simulated wave and observations is presented in Sect \ref{sec:conclusion}.

\section{Numerical Simulation}\label{sec:method}

\begin{figure*}
\begin{interactive}{animation}{Figure1_animation.mp4}
\centering
\includegraphics[width=\textwidth]{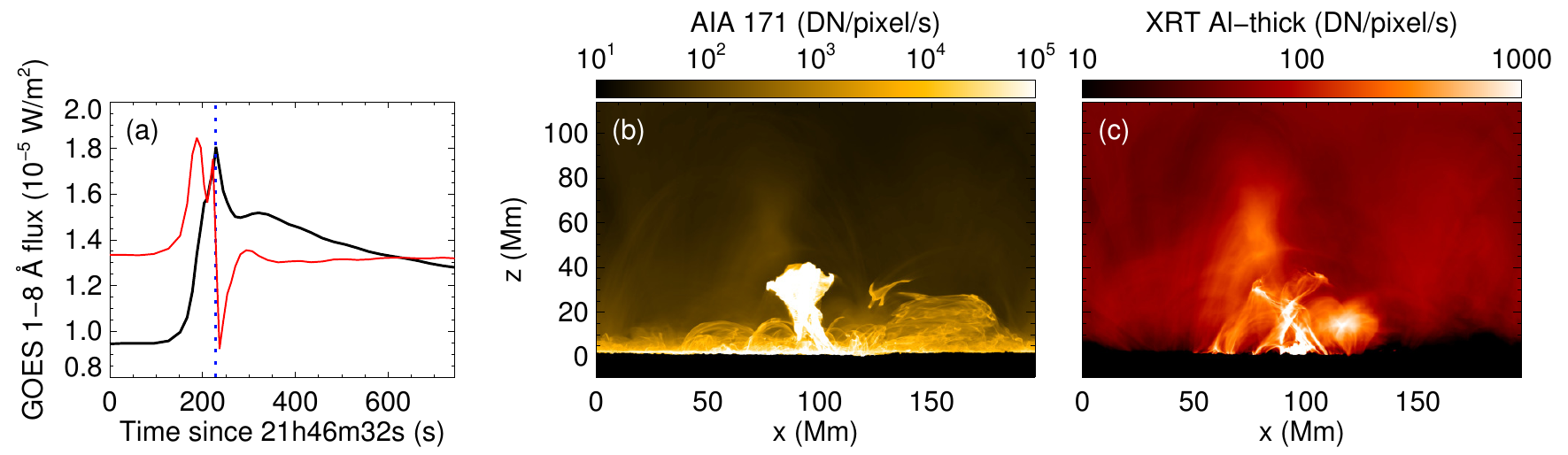}
\end{interactive}
	\caption{Overview \added{of }the flare event \added{simulated by the MURaM code }that triggers the coronal EUV wave. (a) Synthetic GOES-15 1--8\,\AA~flux. The red line shows the temporal derivative of the GOES flux, which is considered as the indicator of the impulsive phase. The blue dotted line indicates the time instance of the image shown on (b) and (c). (b) Synthetic AIA 171\,\AA~image at the flare peak ($t{=}228$\,s). The view point corresponds to observing an event that occurs at the solar limb. (c) Synthetic XRT Al-thick image at the same time and from the same view point as in (b).\added{ The animated version shows an evolution of 743.7 s.}
\newline(An animation of this figure is avaliable.)}\label{fig:overview}
\end{figure*}

We analyze the data from a 3D RMHD simulation spanning from the uppermost convection zone to the corona. The simulation domain is 196.6\,Mm wide in the horizontal directions. The vertical extent is 122.9\,Mm, with the bottom boundary located at 9\,Mm below the photosphere. \added{The domain is resolved by 1024$\times$1024 grid points in the horizontal direction and 1920 grid points in the vertical direction, yielding a spatial resolution of 192\,km and 64\,km in the two directions, respectively.} The simulation was conducted with the MURaM code \citep{Voegler+al:2005,Rempel:2017}, which solves fully compressible MHD equations and implements a sophisticated treatment on the energy balance in the solar atmosphere. The latter is a crucial requirement for making a quantitative comparison between synthesized observables (e.g., AIA images) and real observations.

In this simulation, the magnetic flux bundles generated in a solar convective dynamo \citep{Fan+Fang:2014} are introduced through the bottom boundary. These flux bundles emerge to the photosphere and create sunspots that can reproduce key properties of active regions on the real Sun \citep{Chen+al:2017}. This simulation has a greatly expanded vertical domain that allows the magnetic flux \replaced{emerges }{to emerge }further into the corona. In the full evolution of 48 hours,  the complex and strong active regions formed by the emerging magnetic flux give rise to more than 50 flares in C class and one in M class. A comprehensive analysis on this simulation will be presented in a separate study \citep{Chen+al:2020}. 

In this paper, we focus on a short time period \replaced{that starts from 21h46m32s ($t_{0}$) since the beginning of the simulation}{of 743.7 s from 21h46m32s ($t_{0}$), while the simulation is assumed to start from 00h00m00s} (hereafter, time is shown \deleted{in }relative to $t_{0}$). During this time period, a flare occurs at $t{\approx}200$\,s and generates an evident coronal EUV wave. The synthetic GOES 1--8\,\AA~flux in Figure\,\ref{fig:overview}(a) shows that in the impulsive phase the GOES flux is steeply increased by about $8.5\times10^{-6}$\,W m$^{-2}$ from the pre-flare level. The flare is produced by the eruption of a magnetic flux tube. Figure\,\ref{fig:overview}(b) and \,\ref{fig:overview}(c) displays synthetic AIA 171\,\AA~and XRT Al-thick images at the flare peak ($t{=}228$\,s) from a side view, respectively. The former presents the erupting flux rope, which eventually falls back to the solar surface, and the latter highlights hot plasma giving rise to soft X-ray emission.

\section{Results} \label{sec:result}

\begin{figure*}
\begin{interactive}{animation}{Figure2_animation.mp4}
\centering
\includegraphics[width=\textwidth]{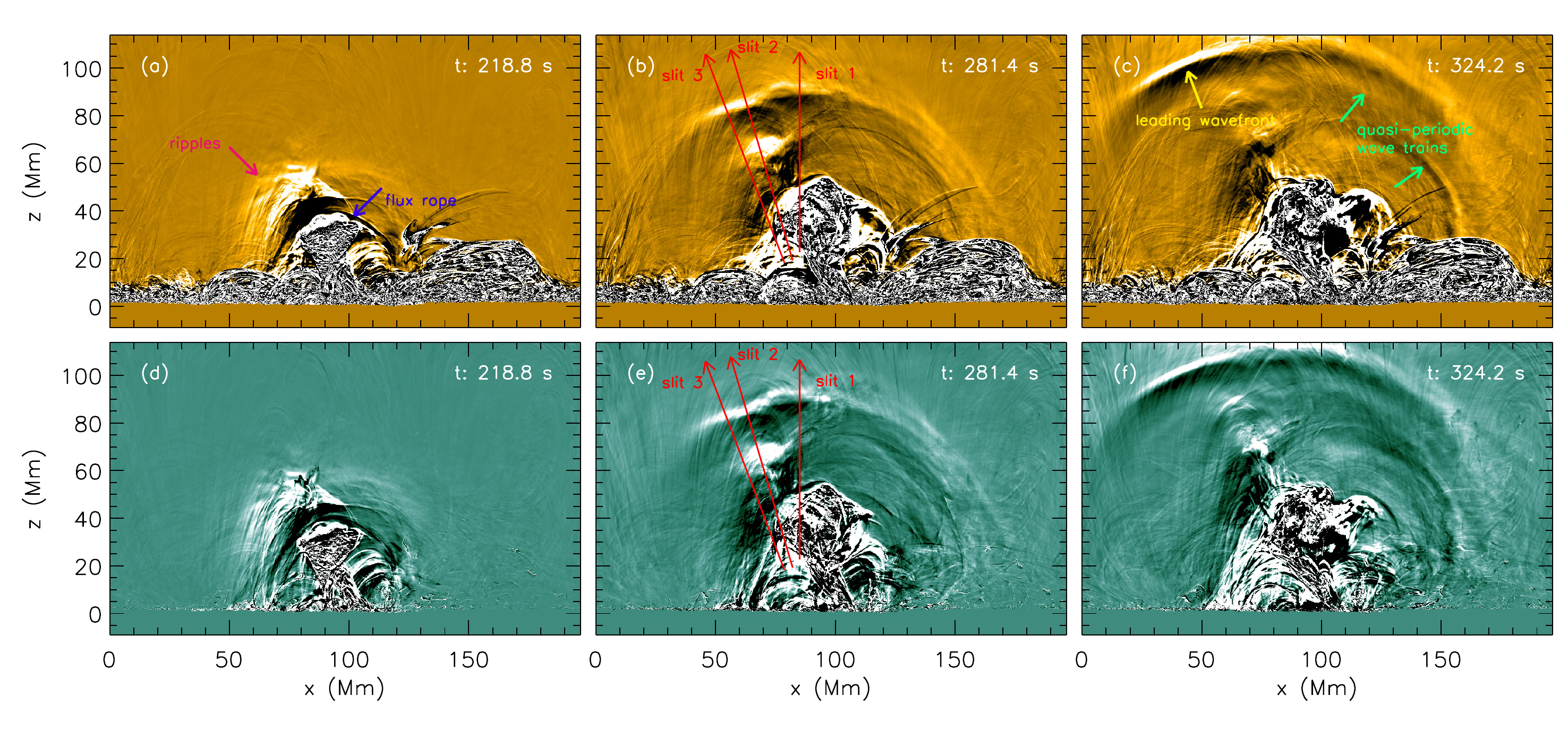}
\end{interactive}
\caption{Running difference of the synthetic AIA 171\,\AA~(top row) and 94\,\AA~(bottom row) images showing the propagation of an EUV wave associated with a flux rope eruption \added{in the simulation.} \added{The pink, blue, yellow, and green arrows indicate features that are described in Section 3.1.} The three red arrows in (b) and (e) refer to the slits used to trace the propagation of the wavefront. \added{The animated version covers a time period from 12.1 to 743.7 s in the simulation.}
\newline(An animation of this figure is avaliable.)}\label{fig:wave17194}
\end{figure*}

\begin{figure*}
\begin{interactive}{animation}{Figure3_animation.mp4}
\centering  
\includegraphics[width=\textwidth]{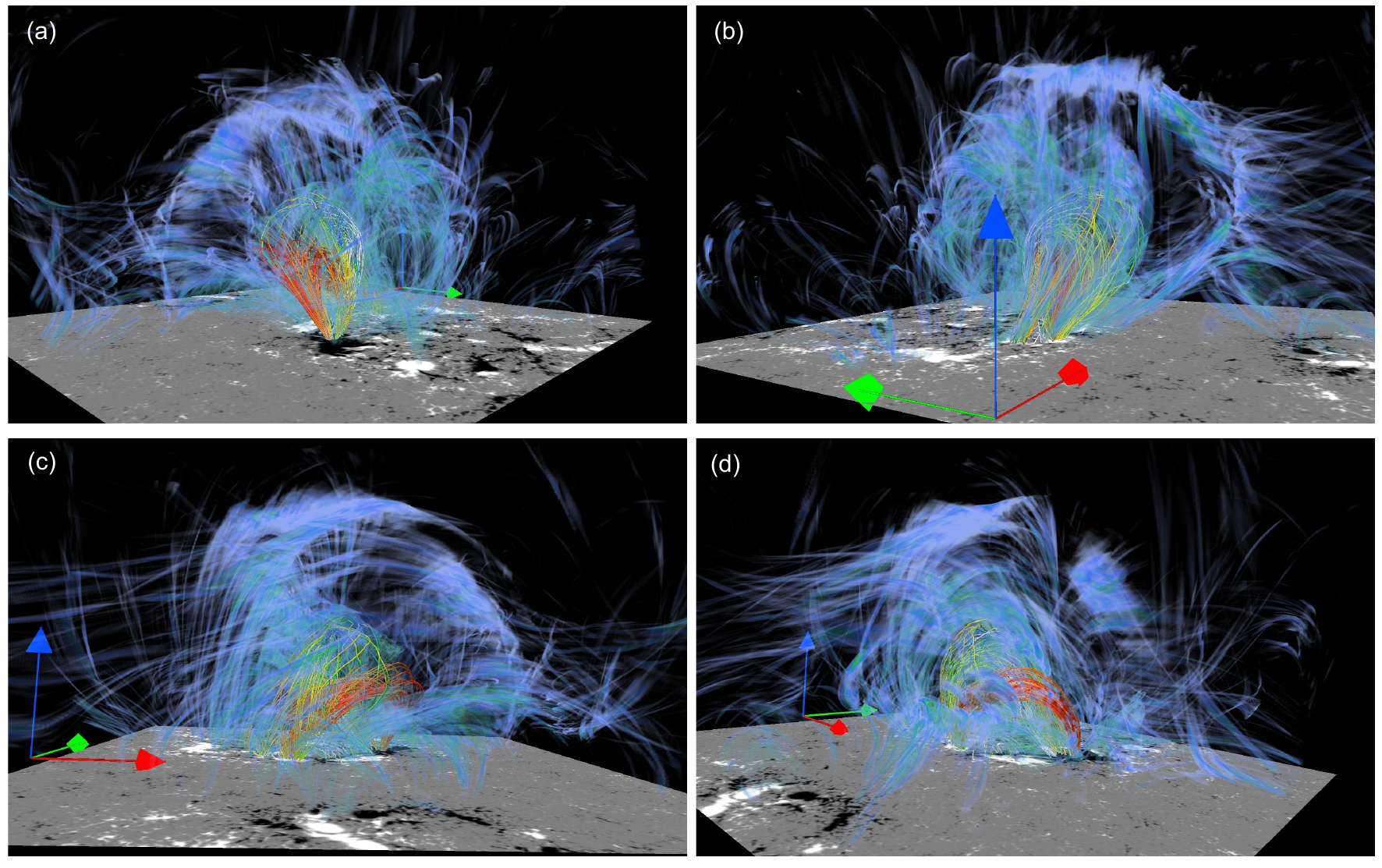}  
\end{interactive}
\caption{The wavefront in the 3D space. This snapshot is at $t{=}281$\,s, which is the same as that for Figure\,\ref{fig:wave17194}. The red, green and blue arrows \replaced{are placed at the origin of the Cartesian coordinate of the simulation and  represent the $x$, $y$, and $z$ axes, respectively.}{identify the origin of the Cartesian coordinate system in the simulation.} The data have been shifted in the $y$ direction (periodic), such that the eruption is placed in the center region of the domain. The blue-colored structure shows the wavefront by the running difference of $n$$_{e}$$^{2}$. The magnetic field lines represent the rising magnetic flux rope, and their color indicates the temperature of plasma there, with white for cool plasma \replaced{in $10^{5}$\,K}{($10^{5}$\,K)} and red for hot plasma \replaced{$10^{7}\,$K}{($10^{7}\,$K)}. \added{The 3D visualization is produced by VAPOR \citep{clyne2007interactive}. }\added{The animated version shows the wavefront from multiple perspectives.}
\newline(An animation of this figure is avaliable.)}\label{fig:wave3d} 
\end{figure*}

\subsection{Morphology of the coronal EUV wave}
The EUV wave can be clearly seen in the running difference of AIA images shown in Figure\,\ref{fig:wave17194}, and the full evolution of the wave is covered by the accompanying animation. 
The wave is ignited during the impulsive phase of the flare. In this phase, instead of a solitary wavefront, many small ripples \added{(pink arrow in Figure\,\ref{fig:wave17194}(a))} are generated at multiple locations surrounding the rising flux rope \added{(blue arrow in Figure\,\ref{fig:wave17194}(a)).} As these ripples propagate outward from the flare site, a large-scale leading wavefront \added{(yellow arrow in Figure\,\ref{fig:wave17194}(c))} is formed. The leading wavefront appears to be a smooth semicircle and sweeps the entire domain. The eruption of the flux rope eventually stops near $z{=}50$\,Mm. However, it can be seen in the animation of Figure\,\ref{fig:wave17194} that the flux rope continues to trigger small ripples that propagate outward. When the apex of the leading wavefront reaches the top of the domain, several weaker wavefronts are formed within the bright leading wavefront and ahead of the flux rope. These wavefronts appear to be very similar to the quasi-periodic wave trains \added{(green arrows in Figure\,\ref{fig:wave17194}(c))} found by \citet{2012ApJ...753...52L}. The top boundary of the domain allows outflows but is not \replaced{a}{perfectly} non-reflective, thus after $t=324$\,s when the apex of the leading wavefront reaches the top of the domain, \replaced{it starts to be reflected.}{a reflection can be observed.} The evolution afterward no longer represents the situation on the real Sun, and hence is excluded from the quantitative analysis. Nonetheless, the reflection \deleted{is a hint on the wave nature of the EUV wave, and }suggests that \replaced{it}{the EUV wave} retains sufficient kinetic energy to propagate farther if not confined by the simulation boundary.

The simulation also provides the opportunity to analyze the wave in the 3D space. For this purpose, we visualize the wavefront by the running difference of electron number density squared ($n_{e}^{2}$), which corresponds to the total emission measure spanning all temperatures. 

Figure\,\ref{fig:wave3d} presents the wavefront in the 3D space and magnetic field lines that outline the flux rope. The perspective of Figure\,\ref{fig:wave3d}\replaced{(e)}{(c)} roughly corresponds to the side view through the $y$-axis. A panoramic view that includes a temporal evolution is shown by the animation associated to this figure. As we can see in the animation, in the early \added{stage of }evolution, the shape of the wavefront generally follows the shape of the flux rope, which indicates the key role of the flux rope eruption in driving the wave. In the later  \added{stage of }evolution when the wavefront has been detached from the flux rope, it becomes a dome-like and highly anisotropic structure in the 3D space that appears to be very different from EUV waves in previous 3D simulations \citep[e.g.,][]{2009ApJ...705..587C,2012ApJ...750..134D,2020MNRAS.493.4816M}. The highly inhomogeneous wavefront is given rise by the \deleted{the }compression of plasma in coronal magnetic field above the complex active regions in this simulation (as shown in the gray-scale images in Figure\,\ref{fig:wave3d}).  A recent multi-perspective observation of coronal EUV wave by \citet{2020SoPh..295..141F} also found the inhomogeneity in the reconstructed wavefront.  

The quasi-periodic wave trains are not visible in the 3D view of the wave event. \replaced{This is because full 3D snapshots of the simulation are stored at a cadence of 2000 iterations which is 10 times lower than that for the synthetic AIA images. The weak signal of the quasi-periodic wave trains is heavily smeared by the running difference of the low cadence data.}{This may be because full 3D snapshots of the simulation are stored at a cadence of 2000 iterations which is 10 times lower than that for the synthetic AIA images. The background corona is very dynamic and undergoes rapid changes during the flare. Therefore, the low amplitude quasi-periodic wave trains are more likely to be contaminated by the changes of the background corona, when taking a running difference with the low cadence data. By comparison, the line-of-sight superimposition of the synthetic 2D AIA images can help to improve the signal-to-noise ratio of the wave trains. However, in the 3D space it becomes more challenging to identify the wave trains.}
It would be intriguing to investigate the behavior of the quasi-periodic wave trains in the 3D space, which will be carried out in a following project with a rerun of the simulation to output 3D data with a sufficiently high cadence.  

\subsection{Kinematics of the coronal EUV wave}

\begin{figure*}
\centering
\includegraphics[width=\textwidth]{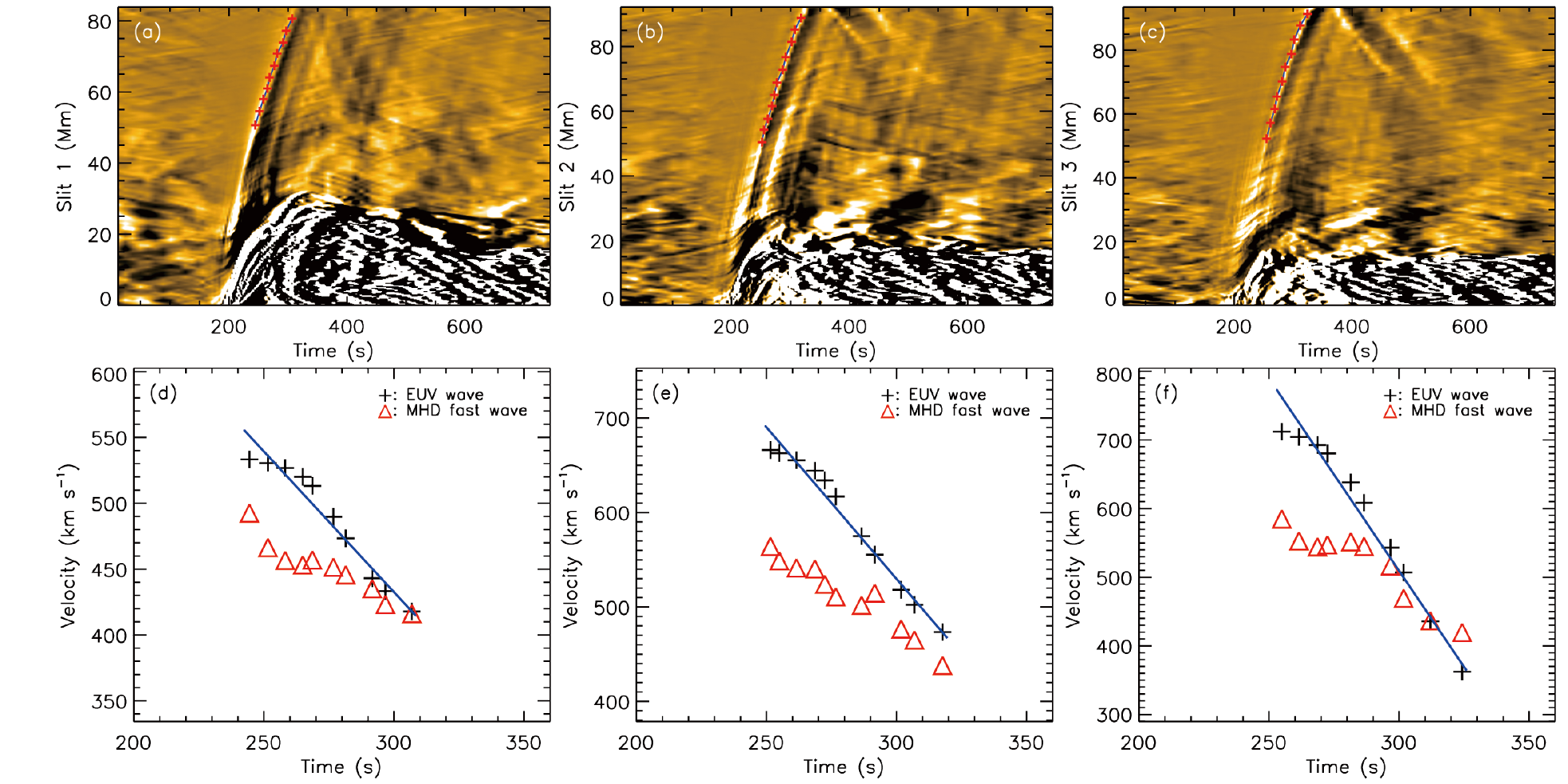}
\caption{Time-distance diagrams and the propagation speeds of the wavefront measured along the three slits. \replaced{The top, middle, and bottom rows are for slits 1, 2, and 3, respectively. In the left column, the horizontal axis represents the sequence number of the simulation snapshots, while in the middle column, it is converted to the real physical time. The positions of the wavefront are marked with red asterisks. The black crosses in the right column show the propagation speed of the wavefront as a function of time, while the red triangles show the local fast magnetoacoustic wave speed.}{The left, middle, and right columns are for slits 1, 2, and 3, respectively. The horizontal axis represents the real physical time. The positions of the wavefront are marked with red crosses in the top row. The black crosses in the bottom row show the propagation speed of the wavefront as a function of time, while the red triangles show the local fast magnetoacoustic wave speed.} The blue lines show linear fittings of the speed of the wavefront.}\label{fig:timedistance}
\end{figure*}

To study the kinematics of the leading wavefront, we select three different points at the wavefront in the running difference of the synthetic AIA 171\,\AA~image at $t=281.4$\,s. At each point, we draw a slit normal to the tangent of the wavefront, which represents the propagation direction of the wavefront (see the three red arrows in Figure\,\ref{fig:wave17194}(b) and\,\ref{fig:wave17194}(e). Because the wavefront remains in a relatively symmetric shape during the propagation, which implies that the kinematics on both sides of the apex are similar, we place all three slits on the left side of the apex for a higher contrast.

We extract along the three slits the intensity of AIA 171\,\AA~running difference images from a time series of 100 snapshots \added{(with a cadence of 13\,s before and after the flare and 2.2\,s in the impulsive phase)}.\deleted{ This creates the time-distance diagrams shown in the left column of Figure\,\ref{fig:timedistance}. We note that, because the time steps in the simulation become significantly smaller during the flare, the time cadences of the snapshots are not uniform (about 13\,s before and after the flare and 2.2\,s in the impulsive phase).} \replaced{In the middle column of Figure\,\ref{fig:timedistance} we also present the same data with the temporal axes shown in physical time.}{This yields the time-distance diagrams shown in the top row of Figure\,\ref{fig:timedistance}.} \explain{We merged two paragraphs of the original manuscript into one in the revised manuscript.}We can discern many short ridges in low contrast near the head of each slits between $t{=}160$ and 200\,s, and some of them could reach much longer distance in the diagram. These correspond to the small ripples triggered by the flux rope eruption during the impulsive phase of the flare as we have seen in Figure\,\ref{fig:wave17194}. However, their low contrast means that it would be almost impossible to identify them in real observations.

The {\it observable} leading wavefront can be clearly identified in the time-distance diagrams as a long and narrow ridge that is significantly brighter than the surroundings. For each slit, we first extract the positions of the wavefront, as marked by the red \replaced{asterisks in the left column}{crosses in the top row} of Figure\,\ref{fig:timedistance}. Then, we perform a cubic Lagrangian interpolation on the curve of the positions (as a function of time), and the derivative of this curve yields the propagation speed of the wavefront. 

The derived speeds along the three slits are plotted as black crosses in Figure\,\ref{fig:timedistance} \replaced{(c), (f), and (i)}{(d), (e), and (f)}, respectively. The blue lines show linear fittings to the propagation speeds. In each direction, the EUV wave decelerates at a roughly constant rate when it propagates farther away from the source region. The deceleration can also be clearly seen in the time-distance diagrams shown in the \replaced{middle column}{top row} of Figure\,\ref{fig:timedistance}, as the slope of the wavefront gradually decreases. When comparing the results among the different slits, we find that the flank of the wavefront (slit 3) has a higher propagation speed in the early stage and a larger deceleration rate afterwards than the apex.

\subsection{The shock nature of the coronal EUV wave}

\begin{figure*}
\centering
\includegraphics[width=\textwidth]{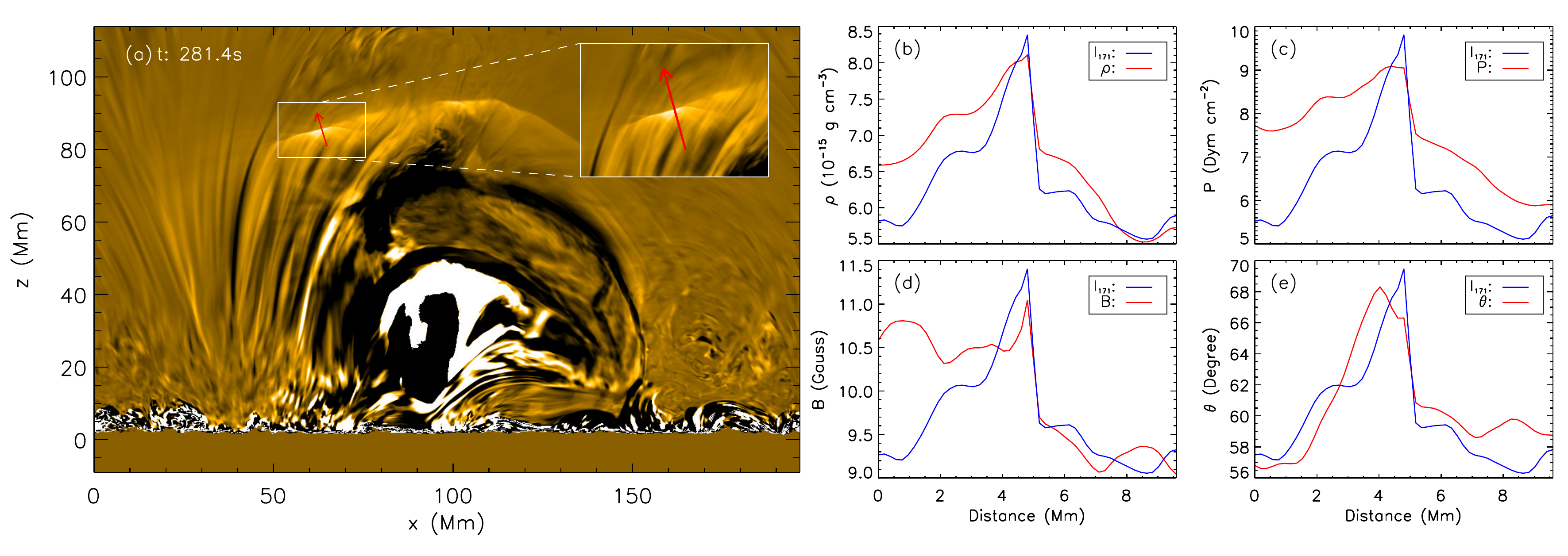}
\caption{Changes of physical quantities across the wavefront. (a) Running difference of synthetic 171\,\AA~image for a 2D vertical slice cutting through the apex of the wavefront. The red arrow indicates the normal direction of the wavefront, a slit along which the physical quantities are displayed in (b)--(e). Variations of the plasma density (b), pressure (c), magnetic field strength (d), and the angle between the magnetic field and the normal direction of the wavefront (e) along the arrow, with the smaller distance on the downstream and larger distance on the upstream. The wavefront is at about 5.5\,Mm. The blue lines in (b)-(e) are identical and plot the AIA 171\,\AA~intensity. It is important to note that all the curves show the original physical quantities but not their running differences.}\label{fig:shock}
\end{figure*}

It is straightforward to clarify in this simulation whether the wavefront is sub- or super-Alfv\'enic, which is also an important question for EUV waves observed on the real Sun.
We calculate the speed of fast magnetoacoustic waves ($v_{\rm f}$) in the simulation domain and plot $v_{\rm f}$ at the location the wavefront passes in the \replaced{right column}{bottom row} of Figure\,\ref{fig:timedistance}. During the early propagation (i.e., before $t{=}280$\,s), the EUV wave in our simulation is clearly a shock with an average Mach number (in relative to $v_{\rm f}$) of about 1.2. However, as the wave travels farther, it gradually degenerates into a fast magnetoacoustic wave.

We further check the changes of plasma properties and magnetic field across the wavefront. For this purpose, we cut through the apex of the dome-shaped wavefront in the 3D space with a vertical slab that is parallel to the $x$-$z$ plane and has an extent of 10\,pixels in the $y$-direction. Then we average the physical quantities over the $y$-direction of the slab, which yields a 2D slice.

We synthesize AIA 171\,\AA~images for this 2D slice and display their running difference at $t{=}281.4$\,s in Figure\,\ref{fig:shock}(a). The wavefront can be clearly identified, and the interface edge (i.e., the shock front) seen in this vertical cut appears to be much sharper than that shown in Figure\,\ref{fig:wave17194}(b), which suffers from the integration along the line-of-sight. The red arrow indicates the normal direction of the wavefront (shock) and marks a segment of 9.7\,Mm. We plot in Figure\,\ref{fig:shock}(b) -\,\ref{fig:shock}(e) the changes of plasma properties and magnetic field along this arrow, with the downstream (shocked medium) on the left of the shock front and upstream (undisturbed medium) on the right. The running difference of AIA images along the arrow is also plotted in these four panels to mark the exact position of the wavefront.

At the wavefront the plasma density (Figure\,\ref{fig:shock}(b)) is increased\footnote{Viewed from the upstream to the downstream (from the right to the left of the plot).} by a factor of 1.2. The pressure (Figure\,\ref{fig:shock}(c)), which also represents the internal energy of the plasma, shows a similar behavior along the arrow and jumps for almost the same factor at the wavefront. This implies that the temperature does not significantly change across the wavefront. It is because the temperature is affected by many physical processes in the corona, and in particular the thermal conduction. 

We present the change of magnetic field strength in Figure\,\ref{fig:shock}(d). The change of the direction of the magnetic field is illustrated by the angle ($\theta$) between the magnetic vector and the normal vector of the wavefront, as shown in Figure\,\ref{fig:shock}(e). There is an abrupt enhancement of the magnetic field strength at the wavefront. 
Meanwhile, the angle also increases accordingly, which means that the tangential component of the magnetic field becomes stronger after the EUV wave has passed by. All the above variations of physical quantities indicate that the leading wavefront is a fast-mode shock.

\section{Discussion and Conclusion} \label{sec:conclusion}
\subsection{Summary of results}
To explore the physical nature of EUV waves, we have analyzed an EUV wave event that occurs during a flare in a comprehensive RMHD simulation of the formation of solar active region through magnetic flux emergence. The EUV wave can be clearly identified from the running difference of synthetic AIA images, as was commonly detected in real observations. The main results are summarized as follows.
\begin{enumerate}
\item The first signature of the wave is many small ripples that are triggered by the lift of the magnetic flux rope during the impulsive phase of the flare. The most prominent wavefront that would be considered as the leading wavefront appears about only 50\,s later and propagates through the entire domain in a semi-circular shape.

\item We find quasi-periodic wave trains from both synthetic AIA images (Figure\,\ref{fig:wave17194}) and time-distance diagrams (Figure\,\ref{fig:timedistance}). They can be seen as secondary wavefronts behind the leading wavefront and ahead of the front edge of the erupting flux rope, and exhibit a period of about 30\,s.

\item The leading wavefront is a fast-mode shock with an initial Mach number (in relative to fast magnetoacoustic wave) of about 1.2. The leading wavefront gradually decelerates to a speed that is similar to the local fast magnetoacoustic wave. 

\item The abrupt increases of the plasma density, pressure, and tangential magnetic field at the sharp wavefront are well resolved and consolidate the (fast-mode) shock nature of the leading wavefront. The change of physical parameters across the wavefront provides a quantitative diagnostic on wave-affected coronal plasma.
\end{enumerate}

\subsection{Large-scale EUV Wave Driven by a Confined Eruption}
The EUV wave in our simulation is clearly a piston-driven MHD shock, which is one of the earliest proposed interpretations \citep[][and references therein]{2015LRSP...12....3W}. Moreover, the wavefront is an integral component of the complex dynamics driven by the eruption of the flux rope, which has also been extensively studied in numerical simulations \citep[e.g., ][]{2002ApJ...572L..99C,2005ApJ...622.1202C,2009ApJ...705..587C,2012ApJ...750..134D,10.1093/mnras/stz2576,2020MNRAS.493.4816M}.

A unique feature of the event we study is that the eruption of the flux rope fails to become a coronal mass ejection (CME). Nevertheless, a large-scale shock/wave is generated by this eruption, and it clearly has a potential to propagate much farther if not limited by the size of the simulation domain. In observations, EUV waves are found to be closely related with CMEs \citep{2005ApJ...631..604C,2006ApJ...641L.153C}. The event we study suggests that confined eruption can drive large-scale waves, as long as sufficient energy is released to the wave, and this can help to understand EUV waves observed in flares without CMEs.

In the unified picture of the EUV waves, in addition to the shock/wave component that can be generated by one strong piston push, a non-wave component is given rise by compression of the plasma outside the leading edge of the expanding CME \citep[see e.g., ][]{2005ApJ...622.1202C,2009ApJ...705..587C,2012ApJ...750..134D}. The non-wave component is absent in the event analyzed in this study. The primary reason is that the flux rope stops rising at the height of about 50\,Mm, and does not provide a persistent push that can compress the plasma in front of the flux rope.

\subsection{Quasi-periodic Wave Trains}

More detailed features of EUV waves revealed by recent observations pose new challenges to numerical models. Our simulation self-consistently reproduces quasi-periodic wave trains within a large-scale EUV wave, as found in the AIA observations \citep[e.g., ][]{2012ApJ...753...52L,2013A&A...554A.144Y,2014A&A...569A..12N,2018ApJ...858L...1Z,2019ApJ...873...22S}. 

Previously numerical simulations of the quasi-periodic wave trains usually employed idealized magnetic configurations and/or artifactual triggers \citep[e.g., ][]{2011ApJ...740L..33O,2013A&A...560A..97P,2016ApJ...823..150T}. On the other hand, large-scale simulations that consider a more realistic setup and coronal physics \citep{2009ApJ...705..587C,2012ApJ...750..134D} were not done with a sufficient spatial resolution to resolve the quasi-periodic wave trains. The simulation we presented in this paper reproduces quasi-periodic wave trains as a component of a large-scale coronal EUV wave that spontaneously occurs in a complex and solar-like active region.

The event in our simulation resembles the general scenario proposed by \citet{2012ApJ...753...52L}, except for the structures inside the CME front. They both follow a process that the eruption of a flux rope (expansion of a CME) drives a leading wavefront that decouples from the driver and multiple periodic wavefronts propagating within the leading wavefront and ahead of the flux rope (CME). 

The wave trains in our simulation exhibit a period of about 30\,s, which is a few times smaller than those seen in large scale events (e.g., 128\,s reported by \citet{2012ApJ...753...52L} and 163\,s reported by \citet{2019ApJ...873...22S}), but is similar to the 45\,s period observed by \citet{2019ApJ...871L...2M}. The periodicity is suggested to originate from the quasi-periodic pulsations in flares \citep{2018SSRv..214...45M}. Therefore, we suspect that the difference in periodicity might be due to the significant difference between the spatial scales of the source region that powers these cyclic behaviors. In order to substantiate the essential cause of the periodicity, it is necessary to investigate the 3D data with a much higher time cadence and preferably with a higher spatial resolution, which would require at least a rerun of the simulation with high cadence output (with improved resolution if resources would permit). This will be carried out in a following project.

To conclude, the RMHD \replaced{simulations}{simulation} can reproduce a comprehensive process of a solar flare that simultaneously drives a large-scale coronal EUV wave and quasi-periodic wave trains. The coronal EUV wave is a fast-mode MHD shock that decays to a fast magnetoacoustic wave. The overall picture is in line with the scenario deduced from decades of theoretical and observational studies of this phenomenon. Last but not least, the high resolution 3D RMHD simulation also indicates that many fined structures and features of coronal EUV wave in the 3D space have not been fully revealed by current remote-sensing observations.

\acknowledgements
The authors thanks the anonymous referee for helpful suggestions that improve the clarity of this paper. FC acknowledges the support of Advanced Study Program (ASP) postdoctoral fellow of NCAR. The authors would like to acknowledge high-performance computing support from Cheyenne (doi:10.5065/D6RX99HX) provided by NCAR's Computational and Information Systems Laboratory, sponsored by the National Science Foundation. This work was also supported by NSFC under grant 11733003.

\bibliography{ms_revision}{}
\bibliographystyle{aasjournal}

\listofchanges
\end{document}